\documentclass[12pt]{article}
\usepackage[superscript]{cite}
\usepackage{CJK}
\usepackage{amssymb}
\usepackage{amsmath}
\usepackage{enumerate}
\usepackage{graphicx}
\usepackage{amsfonts}
\usepackage{url}
\usepackage{bm}
\usepackage{pifont}
\usepackage{epsfig,subfigure,dsfont,amsthm,amsbsy,mathrsfs,amscd}
\usepackage{graphicx}
\usepackage{subfigure}

\def\be{\begin{equation}}
\def\ee{\end{equation}}
\def\ba{\begin{array}}
\def\ea{\end{array}}
\input amssym.def

\newcommand\btd{\raise 2pt \hbox{$\hat\bigtriangledown$}\hskip 1.5pt}
\newcommand\bt{\raise 2pt \hbox{$\bigtriangledown$}\hskip 1.5pt}
\begin{document}

\title{Entropic Uncertainty Principle and Information Exclusion Principle
for multiple measurements in the presence of quantum memory}
\author{Jun Zhang, Yang Zhang, Chang-shui Yu$^\ast$ \\
{\small School of Physics and Optoelectronic Technology, Dalian University
of Technology }\\
{\small Dalian 116024, P. R. China}\\
}
\date{}
\maketitle

\centerline{$^\ast$ Correspondence to: quaninformation@sina.com;
ycs@dlut.edu.cn} \bigskip
\begin{abstract}
The Heisenberg uncertainty principle shows that no one can specify the
values of the non-commuting canonically conjugated variables simultaneously.
However, the uncertainty relation is usually applied to two incompatible
measurements. We present tighter bounds on both entropic uncertainty relation
and information exclusion principle for multiple measurements in the
presence of quantum memory. As applications, three incompatible measurements
on Werner state and Horodecki's bound entangled state are investigated in
details.
\end{abstract}

In quantum mechanics, there is generally an irreducible lower bound on the
uncertainty in the outcomes of simultaneous measurements of noncommuting
observables, i.e., the uncertainty principle which dates back to Heisenberg%
\cite{HUP} , illustrates the the difference between classical and quantum
world and forms the basis of the indeterminacy of quantum mechanics. The
Heisenberg uncertainty principle originally came from a thought experiment
about the measurements of the position and the momentum and later was
generalized by Kennard\cite{kennard} and Robertson \cite{RHUP} to arbitrary
observables $X$ and $Y$ with a strict mathematical formulation $\Delta \hat{X%
}\Delta \hat{Y}\geqslant \frac{1}{2}\left\vert \left\langle \Psi \right\vert
[\hat{X},\hat{Y}]\left\vert \psi \right\rangle \right\vert$ where $\left(
\Delta \hat{X}\right) ^{2}=\left\vert \left\langle \psi \right\vert \left(
\hat{X}-\left\langle \hat{X}\right\rangle \right) ^{2}\left\vert \Psi
\right\rangle \right\vert $ represents the variance and $[\hat{X},\hat{Y}]=%
\hat{X}\hat{Y}-\hat{Y}\hat{X}$ stands for the commutator. However, the standard deviation in 
Robertson's relation is not always a suitable measure of uncertainty \cite{Wehner,inade} . In addition, even though Robertson's relation is good when $X$ and $Y$ are canonically conjugate, the
right-hand side (RHS) of Robertson's relation depends on a state $%
\left\vert\psi\right\rangle$, which will provide a trivial bound if $%
\left\vert\psi\right\rangle$ leads to the \textit{zero} expectation value of the commutator. This kind of uncertainty relations has been studied widely in both theory \cite{Ozawa0,Werner,Wernerz} and
experiment\cite{shiyan,shiyan1,shiyan2,shiyan4,shiyan5,shiyan6} .

Instead of standard deviation, Deutsch \cite{Deutsch} quantified uncertainty
in terms of Shannon entropy and derived the entropic uncertainty relation
(EUR) for any pair of observables\cite{1975shannon} . Later Maassen and
Uffink\cite{Uffink} improved Deutsch's job and gave the following tighter
entropic uncertainty relations:
\begin{equation}
H(X)+H(Y)\geqslant -\log c,
\end{equation}%
where $H(X)$ ($H(Y)$) is the Shannon entropy of measurement outcomes when a
measurement of observable $X$ ($Y$) is performed on a state $\rho $, and $%
c=\max_{i,j}\left\vert \left\langle x_{i}|y_{j}\right\rangle \right\vert
^{2} $ quantifies the complementarity of the non-degenerate observables $X$
and $Y $ with $\left\vert x_{i}\right\rangle ,\left\vert y_{j}\right\rangle $
denoting their eigenvectors, respectively. It is obvious that the bound in
Eq. (1) is state-independent. Hall extended the EUR given by Eq. (1) and
presented an information exclusion principle  which bounds accessible
information about a quantum system given by an ensemble of states when two
observables are performed on it\cite{Hall} . The information exclusion principle for two observable $X$
and $Y$ and the ensemble $\mathcal{E}=\{p_{i},\rho _{i}\}$ is given by
\begin{equation}
I\left( X|\mathcal{E}\right) +I\left( Y|\mathcal{E}\right) \leq 2\log d+\log
c,
\end{equation}%
where $d$ is the dimension of measurement and $I\left( X|\mathcal{E}\right) =H\left( X\right) _{\rho
}-\sum_{i}p_{i}H\left( X\right) _{\rho _{i}}$ is accessible information
about ensemble $\mathcal{E}$ with $X$ performed on it. Both bounds in Eqs
(1,2) have been further improved to different extents \cite%
{Grudka,Piani,Bolan} . The information exclusion principle and especially EUR have been studied widely \cite{Wehner,Heisenberg,Ozawa1,nature,science,Heisen} . It
has been found that EUR has interesting applications in various quantum
information processing tasks ( for example, \cite%
{Wehner,nature1,wit,miyao,miyao1} and references therein). In particular,
considering the direct application in quantum key distribution, Berta
\textit{et al.} \cite{nature} generalized EUR (1) to the case in the
presence of memory, that is,
\begin{equation}
H(X|B)+H(Y|B)\geqslant -\log c+H(A|B),  \label{TS}
\end{equation}%
where  $H(X|B)=S\left( \rho _{XB}\right) -S\left( \rho _{B}\right) $ is the conditional von Neumann entropy and $S(\rho)$ is the von Neumann entropy with $%
\rho _{XB}$ denoting the state after $X$ measurement on subsystem \textit{A}
of $\rho _{AB}$ and $\rho _{B}$ denoting the reduced state of $\rho _{XB}$. Similarly information exclusion principle was also generalized to the case of quantum
memory by replacing the classically mixing ensemble $\mathcal{E}$ with a
quantum system \textit{B} \cite{Piani} , that is,
\begin{equation}
I\left( X:B\right) +I\left( Y:B\right) \leq r_{H}-H\left( A|B\right)
\label{IER}
\end{equation}%
with $r_{H}=\log (d^{2}c)$. In particular, we let IEP abbreviate the information exclusion principle with quantum memory implied. However, most of the relevant jobs usually
consider the case of a pair of observables (measurements).

Recently, the uncertainty relations with multiple measurements have
attracted increasing interests. Significant progresses have been made to
seek for the uncertainty relations for more than two observables \cite%
{Spiros,Liu} , even though the uncertainty relations with two observables
can automatically induce the corresponding uncertainty relations with more
than two observables. In fact, among all the relevant researches, one of the
most fundamental question is that the bounds are not tight enough in general
or precisely speaking, are only tight for some particular states. So in this
paper we would like to present the improved EUR and IEP for multiple
measurements in the presence of quantum memory. One will find that our
bounds for EUR and IEP are generally tighter than previous ones and state-independent, in
particular, it can also be easily reduced to the case without quantum
memory. As applications, we investigate three incompatible measurements on
Werner states and Horodecki's bound entangled states in details.

\medskip \noindent\textbf{Results}

\textsf{Entropic uncertainty relation for multiple measurements in the
presence of quantum memory} To begin with, let's consider an uncertainty
game between Alice and Bob similar to Ref. [\cite{Liu}] . Before the game,
Alice and Bob agree on a group of measurements $\{\Pi_i,i=1,2,\cdots,N\}$
with $\left\vert i_\alpha\right\rangle$ denoting $\alpha $th eigenvector of
the $\Pi_i$. Suppose that Bob prepares a bipartite quantum state $\rho_{AB}$
in $(d\otimes d)$ -dimensional Hilbert space and then sends particle \textit{%
A} to Alice. Alice performs one measurement $\Pi_i$ and announces her choice
to Bob. Bob tries to minimize his uncertainty about Alice's measurement
outcomes.

We proceed by deriving our uncertainty relation. To do so, let's rearrange
the measurements $\{\Pi _{i},i=1,2,\cdots ,N\}$ in a new order with $%
\varepsilon $ denoting the new order. So $\Pi _{\varepsilon _{i}}$ can be
understood as \textit{i}th measurement in the $\varepsilon $ order.
Similarly, the $\alpha $th eigenvector of $\Pi _{\varepsilon _{i}}$ can be
written as $\left\vert \varepsilon _{i}^{\alpha }\right\rangle $. With these
notations, we arrive at the following EUR for the above game in the presence
of quantum memory (Proof given in Methods):
\begin{equation}
\sum\limits_{i=1}^{N}H(\Pi _{i}|B)\geqslant \mathcal{L}_{1}=(N-1)H(A|B)+%
\max_{\varepsilon }\left\{ \ell _{\varepsilon }^{U}\right\} ,  \label{z1}
\end{equation}%
where
\begin{equation}
\ell _{\varepsilon }^{U}=-\sum\limits_{\alpha _{N}}p_{\varepsilon
_{N}^{\alpha _{N}}}\log \sum\limits_{\alpha _{k},N\geqslant k>1}\max_{\alpha
_{1}}\prod\limits_{n=1}^{N-1}\left\vert \left\langle \varepsilon
_{n}^{\alpha _{n}}|\varepsilon _{n+1}^{\alpha _{n+1}}\right\rangle
\right\vert ^{2},
\end{equation}%
with $p_{\varepsilon _{N}^{\alpha }}=Tr\left( \left\vert \varepsilon
_{N}^{\alpha }\right\rangle \left\langle \varepsilon _{N}^{\alpha
}\right\vert \otimes \mathbf{I}\right) \rho _{AB}$. One will find that the
left-hand side (LHS) of Eq. (\ref{z1}) quantifies the total uncertainty
about the measurement outcomes, whilst the right-hand side (RHS) of Eq. (\ref%
{z1}) includes two terms. The first term $H(A|B)$ depends on the initial
state and can describe the effects of entanglement on the EUR. With the
entanglement of $\rho _{AB}$ increasing, the RHS of Eq. (\ref{z1}) could be
negative, but RHS is never negative. At this moment, Eq. (\ref{z1}) will
reduce to a trivial form $\sum\nolimits_{i=1}^{N}H(\Pi _{i}|B)\geqslant 0$.
The second term $\ell _{\varepsilon }^{U}$ depends on the sequence of
observables, the overlap of the projective measurements and the last
observable's probability distribution, it describes the measurement
incompatibility.

When only two measurements $\Pi _{1}$ and $\Pi _{2}$ are considered, by a
simple substitution, our EUR Eq. (\ref{z1}) becomes
\begin{equation}
H(\Pi _{1}|B)+H(\Pi _{2}|B)\geqslant H(A|B)+C_{12},  \label{TO}
\end{equation}%
where $C_{ij}=\max \left\{ C_{ij},C_{ji}\right\} $ with $C_{ij}=-\sum%
\limits_{\alpha _{j}}p_{j}^{\alpha _{j}}\log \max_{\alpha _{i}}\left\vert
\left\langle \alpha _{i}|\alpha _{j}\right\rangle \right\vert ^{2}$. It is
easy to find that this EUR is just consistent with the tight state-dependent
bound of EUR improved by Coles\cite{Piani} . If the state $\rho _{AB}$ is
pure, $H(\Pi _{i}|B)=H(\Pi _{i})-H(B)$ and $H(\rho _{A})=H(\rho _{B})$ \cite{nielsen} . So
the uncertainty relation with quantum memory for pure states $\rho _{AB}$
can be given by
\begin{equation}
\sum\limits_{i=1}^{N}H(\Pi _{i}|B)\geqslant H(A|B)+\max_{\varepsilon
}\left\{ \ell _{\varepsilon }^{U}\right\} .  \label{pure1}
\end{equation}%
Our EUR can be easily reduced to the case without quantum memory. To do so,
we substitute $\rho _{AB}=\rho \otimes \rho _{a}$ into Eq. (\ref{z1}), we
can immediately obtain the EUR for the state $\rho $ without quantum memory as
\begin{equation}
\sum\limits_{i=1}^{N}H(\Pi _{i})\geqslant (N-1)H(\rho )+\max_{\varepsilon
}\left\{ \ell _{\varepsilon }^{U}\right\} .
\end{equation}

It is obvious that the probability distribution in all EUR is a function of
the initial state. In order to eliminate the state-dependency, we will take
maximum over $\alpha _{N}$ of $\Pi _{\varepsilon _{N}}$, so $\ell
_{\varepsilon }^{U}$ in the second term becomes
\begin{eqnarray}
\ell _{\varepsilon }^{U} &=&-\sum\limits_{\alpha _{N}}p_{\varepsilon
_{N}^{\alpha _{N}}}\log \sum\limits_{\alpha _{k},N\geqslant k>1}\max_{\alpha
_{1}}\prod\limits_{n=1}^{N-1}\left\vert \left\langle \varepsilon
_{n}^{\alpha _{n}}|\varepsilon _{n+1}^{\alpha _{n+1}}\right\rangle
\right\vert ^{2}  \notag \\
&\geqslant &-\max_{\alpha _{N}}\log \sum\limits_{\alpha
_{k},N>k>1}\max_{\alpha _{1}}\prod\limits_{n=1}^{N-1}\left\vert \left\langle
\varepsilon _{n}^{\alpha _{n}}|\varepsilon _{n+1}^{\alpha
_{n+1}}\right\rangle \right\vert ^{2}=\ell _{\varepsilon }^{\tilde{U}}.
\label{wu1}
\end{eqnarray}%
Thus, the EUR independent of state can be rewritten as%
\begin{equation}
\sum\limits_{i=1}^{N}H(\Pi _{i}|B)\geqslant \mathcal{\tilde{B}}%
=(N-1)H(A|B)+\max_{\varepsilon }\left\{ \ell _{\varepsilon }^{\tilde{U}%
}\right\} .  \label{Main2}
\end{equation}

As mentioned above, the uncertainty relations for only two observables
actually automatically provides an intuitive bound. Mathematically, Bob can
always employ Eq. (\ref{TO}) (or Eq. (\ref{TS})) for each possible pairs of
measurements of $\{\Pi _{i},i=1,2,\cdots ,N\}$, and then sum the equations
in all kinds of ways and make a proper average finally, so long as he keeps $%
\sum_{i=1}^{N}H\left( \Pi _{i}|B\right) $ in LHS. Bob has many ways to do so
and finally select the maximal one as the bound. It is formally given by
\begin{equation}
\sum_{i=1}^{N}H\left( \Pi _{i}|B\right) \geq \mathcal{L}_{opt}=\frac{N}{2}%
H\left( A|B\right) +\max_{all\text{ }ways}\mathcal{B}_{ways}^{\prime }.
\label{TA}
\end{equation}%
where $\mathcal{B}_{ways}^{\prime} $ is average value of $C_{ij}$ in Eq. (\ref{TO}) for all potential two-measurement combinations. For example, only one way is present for $N=3$ and there are 7 ways for $N=4$. Eq. (\ref{TA}) has consistent form with Eqs. (\ref{z1}) and (\ref{Main2}),
which also shows the effects of entanglement between \textit{A} and \textit{B%
}. Thus we have shown two approaches to obtaining the EUR. However, one will
see that neither alone can serve as a good bound in a general case. They
depend the set of observables. So the tighter EUR should be summarized by
collecting all the contributions (also including all the possible results
that we don't know) as%
\begin{equation}
\sum_{i=1}^{N}H\left( \Pi _{i}|B\right) \geq \max \left\{ \mathcal{L}_{1},%
\mathcal{L}_{opt},0\right\} .  \label{Total}
\end{equation}%
Similarly, the state-independent EUR can also be obtained easily.

\medskip \textsf{Information exclusion relation for multiple measurements in
the presence of quantum memory}~ The IEP was formulated by Hall. It looks
like a transformation of the uncertainty relation based on the mutual
information $I(A:B)=H(\rho _{A})+H(\rho _{B})-H(\rho _{AB})$. Along the
similar game as EUR, Alice and Bob shared a bipartite quantum system $\rho
_{AB}$. Alice performs projective measurements $\left\{ \Pi _{i}\right\} $
on her particle, and the particle at Bob's hand becomes a quantum register
that can record the relevant information. Thus the accessible information is
bounded by the IEP which is given by Eq. (\ref{IER}). The IEP implies that
the information content of quantum observables can be increased only at the
expense of the information carried by complementary observable. It is just a
little difference from the EUR. In particular, one notes that $%
I(A:B)=H(A)-H(A|B)$. Hence we can substitute this relation into the above
EURs and find the corresponding upper bounds on the mutual information,
i.e., the IEP. Following the completely parallel procedure as EUR, we can
present our IER for multiple observables in the presence of memory as

\begin{equation}
\sum\limits_{i=1}^{N}I(\Pi _{i}:B)\leqslant \mathcal{U}_{1}=\sum%
\limits_{i=1}^{N}H(\Pi _{i})-\mathcal{L}_{1}.
\end{equation}

If we limit only two projective measurements $\Pi _{1}$ and $\Pi _{2}$, the
IEP will reduce to
\begin{equation}
I(\Pi _{1}:B)+I(\Pi _{2}:B)\leqslant H(\Pi _{1})+H(\Pi _{2})-H(A|B)-C_{12}.
\label{liang1}
\end{equation}%
Analogous to EUR, for multiple measurements one can also select any pair of
observables and use the IEP given in Eq. (\ref{liang1}). Thus one will
obtain a series of equations. Keep $\sum\limits_{i=1}^{N}I(\Pi _{i}:B)$ in
the LHS, one will give an upper bound. Considering different combinations of
the observables, one can obtain many upper bounds. We choose the minimal one
as the final upper bound. Hence, such an IEP can be formally given by%
\begin{equation}
\sum\limits_{i=1}^{N}I(\Pi _{i}:B)\leqslant \mathcal{U}_{opt}.  \label{IEPI}
\end{equation}%
Thus the tighter bound for IEP should be written as%
\begin{equation}
\sum\limits_{i=1}^{N}I(\Pi _{i}:B)\leqslant \min \left\{ \mathcal{U}_{1},%
\mathcal{U}_{opt}\right\} .
\end{equation}

Similarly, from Eq. (\ref{IEPI}), one can obtain a state-independent upper
bound denoted by $\mathcal{\tilde{U}}_{opt}$. From Eq. (\ref{Main2}), one
can get the state-independent IEP as
\begin{equation}
\sum\limits_{i=1}^{N}I(\Pi _{i}:B)\leqslant \mathcal{\tilde{U}}_{1}=N\log d-%
\mathcal{\tilde{B}}  \label{IEPU}
\end{equation}%
with $\mathcal{\tilde{B}}$ defined in Eq. (\ref{IEPI}).
The IEP given in Eq. (\ref{IEPU}) is obtained by taking the maximum
probability $p_{\varepsilon _{N}}^{\alpha _{N}}$. Alternatively, we can
employ the concavity of the logarithm to find another bound as
\begin{equation}
\sum\limits_{i=1}^{N}I(\Pi _{i}:B)\leqslant \mathcal{\tilde{U}}%
_{2}=(N-1)\log d-(N-1)H(A|B)+\min_{\varepsilon }\left\{ {\normalsize u}%
_{\varepsilon }^{I}\right\} ,  \label{ourIEP}
\end{equation}%
with
\begin{equation}
{\normalsize u}_{\varepsilon }^{I}=\log \sum\limits_{\alpha _{k},N\geqslant
k>1}\max\limits_{\alpha _{1}}\prod\limits_{n=1}^{N-1}\left\vert \left\langle
\varepsilon _{n}^{\alpha _{n}}|\varepsilon _{n+1}^{\alpha
_{n+1}}\right\rangle \right\vert ^{2}.
\end{equation}%
Summarizing Eq. (\ref{IEPU}) and Eq. (\ref{ourIEP}) as well as $\mathcal{%
\tilde{U}}_{opt}$, one can write the state-independent IEP as%
\begin{equation}
\sum\limits_{i=1}^{N}I(\Pi _{i}:B)\leqslant \min \left\{ \mathcal{\tilde{U}}%
_{1},\mathcal{\tilde{U}}_{2},\mathcal{\tilde{U}}_{opt}\right\} .
\label{totalIEP}
\end{equation}
The necessary derivations of the results in Eq. (\ref{totalIEP}) are given in Methods.

\medskip \textsf{Applications for three projective measurements}~As
applications, we first consider three 2-dimensional observables measured on
the Werner state which is given by \cite{wernerstate}
\begin{equation}
\rho _{AB}=\eta \left\vert \psi ^{\dag}\right\rangle \left\langle \psi
^{\dag}\right\vert +\frac{1-\eta }{4}\mathbf{I},
\end{equation}%
with $\left\vert \psi ^{\dag }\right\rangle =\frac{1}{\sqrt{2}}(\left\vert
00\right\rangle +\left\vert 11\right\rangle )$ the maximally entangled state
and $0\leqslant \eta \leqslant 1.$
\begin{figure*}[tbp]
\begin{center}
\hspace*{-7.5cm} \includegraphics*[width=30cm]{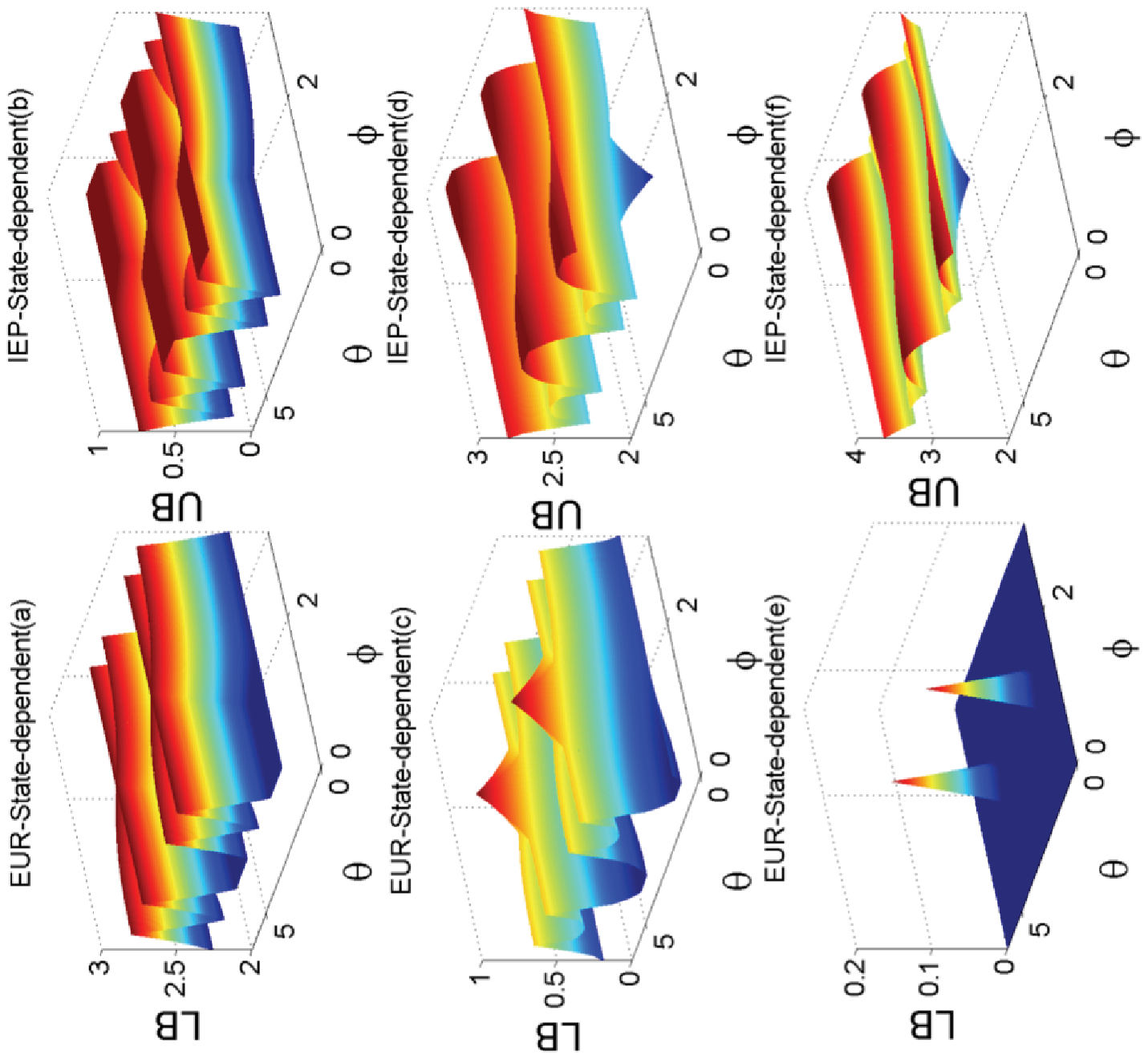}
\end{center}
\caption{(color online) The bounds of entropic uncertainty relation and
information exclusion principle for the three measurements in
two-dimensional space in the presence of quantum memory vs. the azimuthal
angle $\protect\varphi $ and the polar $\protect\theta $ of the first
observable. The left column (a), (c), (e) correspond to the entropic
uncertainty relation and the right column (b), (d), (f) correspond to the
information exclusion principle. From the top to the bottom, the purity $%
\protect\eta $ of Werner state takes $0.2,0.8$ and $0.95$, respectively.}
\end{figure*}
\begin{figure*}[tbp]
\begin{center}
\includegraphics*[width=12cm,height=10cm]{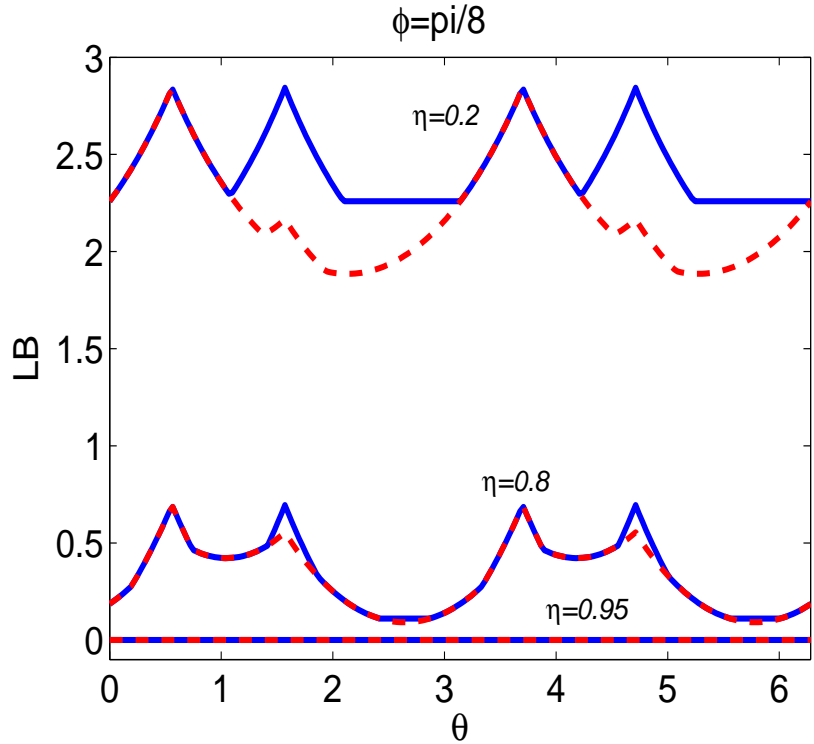}
\end{center}
\caption{(color online) The state-dependent bounds of EUR vs. the polar $\protect\theta $ when the azimuthal
angle $\protect\varphi =\pi/8$ of the first
observable. The blue lines correspond to the state-dependent bound of entropic
uncertainty relation in Eq. (\ref{Total}) while the red dash lines correspond to the previous one in Ref.\cite{Liu} . From the top to the bottom, the purity $\protect\eta $ of Werner state takes $0.2,0.8$ and $0.95$, respectively.}
\end{figure*}
Let ${X}$ denote an observable with the eigenvectors given by
\begin{equation}
X:\{(\cos \frac{\theta }{2},-e^{i\phi }\sin \frac{\theta }{2}),(e^{-i\phi
}\sin \frac{\theta }{2},\cos \frac{\theta }{2})\}.
\end{equation}%
Similarly, we can define the other two observables ${Y}$ and $Z$ as follows:
\begin{eqnarray}
Y &:&\{(\frac{1}{2},\frac{\sqrt{3}}{2}),(\frac{\sqrt{3}}{2},-\frac{1}{2})\},
\\
Z &:&\{(1,0),(0,1)\}.
\end{eqnarray}%
As an example, we only illustrate the state-dependent EUR and IEP. The
bounds of EUR and IEP with various purities $\eta $ of the Werner state are
plotted in Fig. 1 . As we know, if the purity $0\leq \eta \leq 1/3$, the
Werner state is separable. Fig. 1(a) shows that the shape of the bounds of
EUR looks like a double alphabet "X" when the Werner state includes no
entanglement. However, with the purity increasing, the bounds of EUR will
become small due to the generation of entanglement of the Werner state,
which is given in Fig. 1(c). But the crossing point of the alpahbet "X"
reduces slowly. With the purity getting much stronger, the bound of the
entropic uncertainty relation is shown in Fig. 1(e) with $\eta =0.95$. The
crossing points of the double alphabet "X" becomes two peaks. If the purity $%
\eta $ gets stronger and stronger, which means that the entanglement of the
Werner state becomes much larger, the bounds of the EUP will decrease
further until it goes down to $0$. At that moment, the bound is trivial. The
opposite behaviors can be found for the IEP which are illustrated by Fig. 1
(b), (d) and (f). However, one can find that the bounds of IEP is still
acceptable, even though the bounds for EUR could be trivial. While in Fig. 2, we set the azimuthal
angle $\protect\varphi =\pi/8$ of the first observable, the blue lines correspond to the state-dependent bound of entropic
uncertainty relation in Eq. (\ref{Total}) while the red dash lines correspond to the previous one in Ref.\cite{Liu} . One can find that our bound is tighter than previous one.

\begin{figure*}[tbp]
\begin{center}
\hspace*{-7.5cm} \includegraphics*[width=30cm]{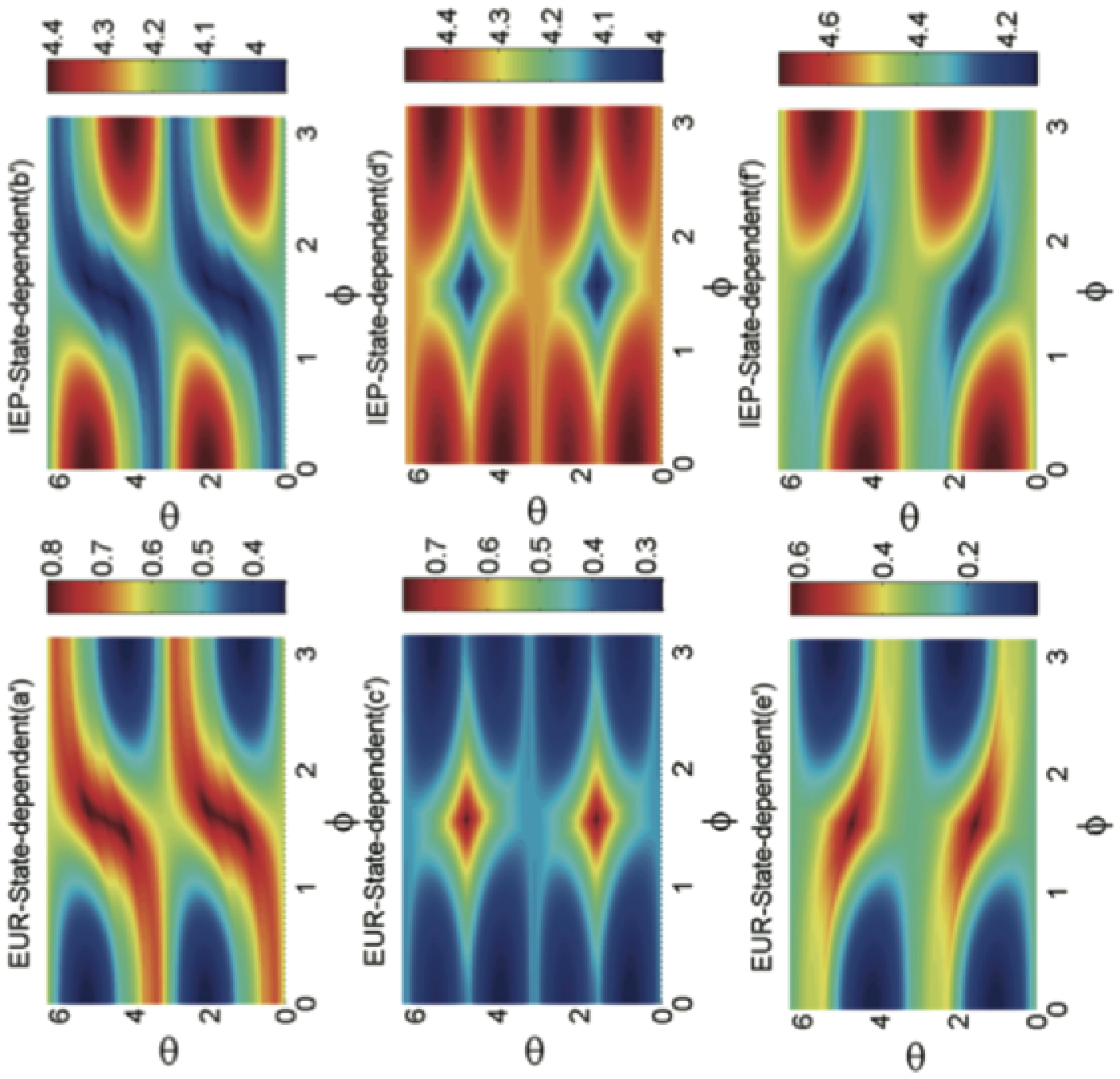}
\end{center}
\caption{(color online) The bounds of entropic uncertainty relation and
information exclusion principle for the three measurements in
three-dimensional space in the presence of quantum memory vs. the azimuthal
angle $\protect\varphi $ and the polar $\protect\theta $ of the first
observable. The left column (a'), (c'), (e') correspond to the entropic
uncertainty relation and the right column (b'), (d'), (f') correspond to the
information exclusion principle. In all cases, $\protect\alpha=0.6 $.}
\end{figure*}
Next, we consider another example with three observables in
three-dimensional Hilbert space. Here the measured state is the Horodecki's
bound entangled state which reads\cite{horodec}
\begin{equation}
\rho _{AB}=\frac{1}{8\alpha +1}\left(
\begin{array}{ccccccccc}
\alpha & 0 & 0 & 0 & \alpha & 0 & 0 & 0 & \alpha \\
0 & \alpha & 0 & 0 & 0 & 0 & 0 & 0 & 0 \\
0 & 0 & \alpha & 0 & 0 & 0 & 0 & 0 & 0 \\
0 & 0 & 0 & \alpha & 0 & 0 & 0 & 0 & 0 \\
\alpha & 0 & 0 & 0 & \alpha & 0 & 0 & 0 & \alpha \\
0 & 0 & 0 & 0 & 0 & \alpha & 0 & 0 & 0 \\
0 & 0 & 0 & 0 & 0 & 0 & \beta & 0 & \gamma \\
0 & 0 & 0 & 0 & 0 & 0 & 0 & \alpha & 0 \\
\alpha & 0 & 0 & 0 & \alpha & 0 & \gamma & 0 & \beta%
\end{array}%
\right) ,
\end{equation}%
with $\beta =\frac{1+\alpha }{2}$ and $\gamma =\frac{\sqrt{1-\alpha ^{2}}}{2}
$. The eigenvectors for the first observable $X$ is supposed to be
\begin{equation}
X:\left\{ (\cos \frac{\theta }{2},-e^{i\phi }\sin \frac{\theta }{2}%
,0),(e^{-i\phi }\sin \frac{\theta }{2},\cos \frac{\theta }{2}%
,0),(0,0,1)\right. .
\end{equation}%
In addition, we randomly generate $3$ groups of observables \{${Y}$, $Z\}$
with the eigenvectors of $Y$ and $Z$ given respectively by
\begin{equation}
\begin{array}{l}
\left\{
\begin{array}{l}
Y:\left\{ {\left( {0.3282,-0.9425,0.0633}\right) ,\left( {%
0.6684,0.1843,-0.7206}\right) ,\left( {0.6675,0.2788,0.6904}\right) }\right\}
\\
Z:\left\{ {\left( {-0.1355,0.4003,-0.9063}\right) ,\left( {%
0.6065,-0.6898,-0.3953}\right) ,\left( {0.7835,0.6032,0.1493}\right) }%
\right\}%
\end{array}%
\right. , \\
\left\{
\begin{array}{l}
Y:\left\{ {\left( {-0.1429,-0.4205,0.8960}\right) ,\left( {%
-0.7427,0.6439,0.1837}\right) ,\left( {-0.6542,-0.6392,-0.4043}\right) }%
\right\} \\
Z:\left\{ {\left( {0.8783,-0.0955,-0.4685}\right) ,\left( {%
0.1058,-0.9168,0.3852}\right) ,\left( {0.4663,0.3879,0.7951}\right) }\right\}%
\end{array}%
\right. , \\
\left\{
\begin{array}{l}
Y:\left\{ {\left( {0.4514,0.6672,-0.5925}\right) ,\left( {%
0.6676,-0.6931,-0.2719}\right) ,\left( {0.5920,0.2728,0.7583}\right) }%
\right\} \\
Z:\left\{ {\left( {-0.8182,0.3974,0.4155}\right) ,\left( {%
-0.2143,-0.8814,0.4210}\right) ,\left( {0.5335,0.2554,0.8063}\right) }%
\right\}%
\end{array}%
\right. .%
\end{array}
\label{san}
\end{equation}%
The bounds of EUR and IEP versus $\theta $ and $\varphi $ are plotted in
Fig. 2. The left column in Fig. 2 corresponds to the lower bounds of EUR and
the right column corresponds to the upper bounds of IEP. Each row
corresponds to one choice of Eq. (\ref{san}). All the figures show the
tightness of our bounds.

\medskip \noindent \textbf{Discussions}

Uncertainty relations are the fundamental features of quantum mechanics and
have wide applications in quantum information processing tasks. We have
considered the EUR and IEP for more than two observables in the presence of
quantum memory and presented tight bounds for them. From our results one can
easily obtain the EUR in the absence of quantum memory. The nontrivial
bounds of EUR and IEP can be determined by the complementary of the
measurements and the entanglement of the composite system. As a consequence,
the nontrivial bounds shed new light on quantum uncertainty.

\medskip \noindent\textbf{Methods}

Before the proof of Eq. (\ref{z1}), we would like first to give a lemma.

\textbf{Lemma} For a bipartite quantum system $\rho _{AB}$ and a group of
measurements $\{\Pi _{i},i=1,2,\cdots ,N\}$ which are performed on the
subsystem $A$, there will have the following relations:
\begin{equation}
\sum\limits_{i=1}^{N}H(\Pi _{i}|B)-NH(A|B)\geqslant S\left( \rho
_{AB}\left\Vert \sum\limits_{\alpha _{k},N\geqslant k\geqslant
1}\prod\limits_{n=1}^{N-1}\left\vert \left\langle \varepsilon _{n}^{\alpha
_{n}}|\varepsilon _{n+1}^{\alpha _{n+1}}\right\rangle \right\vert
^{2}\left\vert \varepsilon _{N}^{\alpha _{N}}\right\rangle \left\langle
\varepsilon _{N}^{\alpha _{N}}\right\vert \otimes \rho _{B}^{\alpha
_{1}}\right. \right) ,
\end{equation}%
with $S\left( \rho _{AB}\left\Vert \cdot \right. \right) $ denoting the
relative entropy.

\textsf{Proof.}~ First, we prove that a pair of the projective measurements $%
\Pi _{1}$ and $\Pi _{2}$ are acted on the inital quantum state, the above
relation hold. That is, for $N=2$, we have
\begin{eqnarray}
&&H(\Pi _{\varepsilon _{1}}|B)-H(A|B)  \notag \\
&=&H\left( \rho _{AB}\left\Vert \Pi _{\varepsilon _{1}}\rho _{AB}\Pi
_{\varepsilon _{1}}\right. \right)  \notag \\
&\geqslant &H\left( \Pi _{\varepsilon _{2}}\rho _{AB}\Pi _{\varepsilon
_{2}}\left\Vert \Pi _{\varepsilon _{2}}\left( \Pi _{\varepsilon _{1}}\rho
_{AB}\Pi _{\varepsilon _{1}}\right) \Pi _{\varepsilon _{2}}\right. \right)
\notag \\
&=&-H(\Pi _{\varepsilon _{2}}B)-Tr\Pi _{\varepsilon _{2}}\rho _{AB}\Pi
_{\varepsilon _{2}}\ln \sum\limits_{\alpha _{1},\alpha _{2}}\left\vert
\left\langle \varepsilon _{2}^{\alpha _{2}}|\varepsilon _{1}^{\alpha
_{1}}\right\rangle \right\vert ^{2}\left\vert \varepsilon _{2}^{\alpha
_{2}}\right\rangle \left\langle \varepsilon _{2}^{\alpha _{2}}\right\vert
\otimes \rho _{B}^{\alpha _{1}}  \notag \\
&=&-H(\Pi _{\varepsilon _{2}}B)-Tr\rho _{AB}\ln \sum\limits_{\alpha
_{1},\alpha _{2}}\left\vert \left\langle \varepsilon _{2}^{\alpha
_{2}}|\varepsilon _{1}^{\alpha _{1}}\right\rangle \right\vert ^{2}\left\vert
\varepsilon _{2}^{\alpha _{2}}\right\rangle \left\langle \varepsilon
_{2}^{\alpha _{2}}\right\vert \otimes \rho _{B}^{\alpha _{1}}+S(\rho
_{B})-S(\rho _{B})+S(\rho _{AB})-S(\rho _{AB})  \notag \\
&=&-H(\Pi _{\varepsilon _{2}}|B)+H(A|B)+H\left( \rho _{AB}\left\Vert
\sum\limits_{\alpha _{1},\alpha _{2}}\left\vert \left\langle \varepsilon
_{2}^{\alpha _{2}}|\varepsilon _{1}^{\alpha _{1}}\right\rangle \right\vert
^{2}\left\vert \varepsilon _{2}^{\alpha _{2}}\right\rangle \left\langle
\varepsilon _{2}^{\alpha _{2}}\right\vert \otimes \rho _{B}^{\alpha
_{1}}\right. \right) .
\end{eqnarray}%
Here the inequality holds because of the adjoint concavity of relative
entropy, i.e., $H\left( \rho ||\sigma \right) \geqslant H\left( \$(\rho
)||\$(\sigma )\right) $ with $\$(\cdot )$ denoting a superoperator. Thus,
for a pair of measurements applied on the subsystem $A$, the following
relation is satisfied:%
\begin{equation}
H(\Pi _{1}|B)+H(\Pi _{2}|B)-2H(A|B)\geqslant H\left( \rho _{AB}\left\Vert
\sum\limits_{\alpha _{1},\alpha _{2}}\left\vert \left\langle \varepsilon
_{2}^{\alpha _{2}}|\varepsilon _{1}^{\alpha _{1}}\right\rangle \right\vert
^{2}\left\vert \varepsilon _{2}^{\alpha _{2}}\right\rangle \left\langle
\varepsilon _{2}^{\alpha _{2}}\right\vert \otimes \rho _{B}^{\alpha
_{1}}\right. \right) .
\end{equation}%
Now, let's assume that when a set of nondegenerate measurements $\{\Pi
_{i},i=1,2,\cdots ,N\}$ are performed on the subsystem $A$, the inequality
hold for the $N$ measurements. Thus, considering the set of measurements $%
\{\Pi _{i},i=1,2,\cdots ,N,N+1\}$, we have
\begin{eqnarray}
&&\sum\limits_{i=1}^{N}H(\Pi _{i}|B)-NH(A|B)  \notag \\
&\geqslant &H\left( \rho _{AB}\left\Vert \sum\limits_{\alpha _{k},N\geqslant
k\geqslant 1}\prod\limits_{n=1}^{N-1}\left\vert \left\langle \varepsilon
_{n}^{\alpha _{n}}|\varepsilon _{n+1}^{\alpha _{n+1}}\right\rangle
\right\vert ^{2}\left\vert \varepsilon _{N}^{\alpha _{N}}\right\rangle
\left\langle \varepsilon _{N}^{\alpha _{N}}\right\vert \otimes \rho
_{B}^{\alpha _{1}}\right. \right)  \notag \\
&\geqslant &H\left( \Pi _{\varepsilon _{N+1}}\rho _{AB}\Pi _{\varepsilon
_{N+1}}\left\Vert \Pi _{\varepsilon _{N+1}}\left( \sum\limits_{\alpha
_{k},N\geqslant k\geqslant 1}\prod\limits_{n=1}^{N-1}\left\vert \left\langle
\varepsilon _{n}^{\alpha _{n}}|\varepsilon _{n+1}^{\alpha
_{n+1}}\right\rangle \right\vert ^{2}\left\vert \varepsilon _{N}^{\alpha
_{N}}\right\rangle \left\langle \varepsilon _{N}^{\alpha _{N}}\right\vert
\otimes \rho _{B}^{\alpha _{1}}\right) \Pi _{\varepsilon _{N+1}}\right.
\right)  \notag \\
&=&H\left( \Pi _{\varepsilon _{N+1}}\rho _{AB}\Pi _{\varepsilon
_{N+1}}\left\Vert \sum\limits_{\alpha _{k},N+1\geqslant k\geqslant
1}\prod\limits_{n=1}^{N}\left\vert \left\langle \varepsilon _{n}^{\alpha
_{n}}|\varepsilon _{n+1}^{\alpha _{n+1}}\right\rangle \right\vert
^{2}\left\vert \varepsilon _{N+1}^{\alpha _{N+1}}\right\rangle \left\langle
\varepsilon _{N+1}^{\alpha _{N+1}}\right\vert \otimes \rho _{B}^{\alpha
_{1}}\right. \right)  \notag \\
&=&-H(\Pi _{\varepsilon _{N+1}}B)-Tr\Pi _{\varepsilon _{N+1}}\rho _{AB}\Pi
_{\varepsilon _{N+1}}\ln \sum\limits_{\alpha _{k},N+1\geqslant k\geqslant
1}\prod\limits_{n=1}^{N}\left\vert \left\langle \varepsilon _{n}^{\alpha
_{n}}|\varepsilon _{n+1}^{\alpha _{n+1}}\right\rangle \right\vert
^{2}\left\vert \varepsilon _{N+1}^{\alpha _{N+1}}\right\rangle \left\langle
\varepsilon _{N+1}^{\alpha _{N+1}}\right\vert \otimes \rho _{B}^{\alpha _{1}}
\notag \\
&=&-H(\Pi _{\varepsilon _{N+1}}B)-Tr\rho _{AB}\ln \sum\limits_{\alpha
_{k},N+1\geqslant k\geqslant 1}\prod\limits_{n=1}^{N}\left\vert \left\langle
\varepsilon _{n}^{\alpha _{n}}|\varepsilon _{n+1}^{\alpha
_{n+1}}\right\rangle \right\vert ^{2}\left\vert \varepsilon _{N+1}^{\alpha
_{N+1}}\right\rangle \left\langle \varepsilon _{N+1}^{\alpha
_{N+1}}\right\vert \otimes \rho _{B}^{\alpha _{1}}  \notag \\
&=&-H(\Pi _{\varepsilon _{N+1}}|B)+H(A|B)+H\left( \rho _{AB}\left\Vert
\sum\limits_{\alpha _{k},N+1\geqslant k\geqslant
1}\prod\limits_{n=1}^{N}\left\vert \left\langle \varepsilon _{n}^{\alpha
_{n}}|\varepsilon _{n+1}^{\alpha _{n+1}}\right\rangle \right\vert
^{2}\left\vert \varepsilon _{N+1}^{\alpha _{N+1}}\right\rangle \left\langle
\varepsilon _{N+1}^{\alpha _{N+1}}\right\vert \otimes \rho _{B}^{\alpha
_{1}}\right. \right) .  \notag \\
&&
\end{eqnarray}%
Rearrange the above inequality, we will find that
\begin{equation}
\sum\limits_{i=1}^{N+1}H(\Pi _{i}|B)-(N+1)H(A|B)\geqslant H\left( \rho
_{AB}\left\Vert \sum\limits_{\alpha _{k},N+1\geqslant k\geqslant
1}\prod\limits_{n=1}^{N}\left\vert \left\langle \varepsilon _{n}^{\alpha
_{n}}|\varepsilon _{n+1}^{\alpha _{n+1}}\right\rangle \right\vert
^{2}\left\vert \varepsilon _{N+1}^{\alpha _{N+1}}\right\rangle \left\langle
\varepsilon _{N+1}^{\alpha _{N+1}}\right\vert \otimes \rho _{B}^{\alpha
_{1}}\right. \right) .
\end{equation}%
During this process, we let the first measurement $\Pi _{\varepsilon _{1}}$
perform on the local system $A$ and use $H\left( \rho _{AB}\left\Vert \Pi
_{\varepsilon _{1}}\rho _{AB}\Pi _{\varepsilon _{1}}\right. \right) =H(\Pi
_{\varepsilon _{1}}|B)-H(A|B)$. In addition, the first and the second
inequalities are satisfied again due to the adjoint concavity of relative
entropy. The proof of the lemma is completed.$\hfill {}\blacksquare $

\textsf{Proof of the Eq. (\ref{z1}).} Using the lemma, the EUR of $N$
measurements can be given as follows.
\begin{eqnarray}
&&\sum\limits_{i=1}^{N-1}H(\Pi _{i}|B)-(N-1)H(A|B)  \notag \\
&\geqslant &H\left( \rho _{AB}\left\Vert \sum\limits_{\alpha
_{k},N-1\geqslant k\geqslant 1}\prod\limits_{n=1}^{N-2}\left\vert
\left\langle \varepsilon _{n}^{\alpha _{n}}|\varepsilon _{n+1}^{\alpha
_{n+1}}\right\rangle \right\vert ^{2}\left\vert \varepsilon _{N-1}^{\alpha
_{N-1}}\right\rangle \left\langle \varepsilon _{N-1}^{\alpha
_{N-1}}\right\vert \otimes \rho _{B}^{\alpha _{1}}\right. \right)   \notag \\
&\geqslant &H\left( \Pi _{\varepsilon _{N}}\rho _{AB}\Pi _{\varepsilon
_{N}}\left\Vert \Pi _{\varepsilon _{N}}\left( \sum\limits_{\alpha
_{k},N-1\geqslant k\geqslant 1}\prod\limits_{n=1}^{N-2}\left\vert
\left\langle \varepsilon _{n}^{\alpha _{n}}|\varepsilon _{n+1}^{\alpha
_{n+1}}\right\rangle \right\vert ^{2}\left\vert \varepsilon _{N-1}^{\alpha
_{N-1}}\right\rangle \left\langle \varepsilon _{N-1}^{\alpha
_{N-1}}\right\vert \otimes \rho _{B}^{\alpha _{1}}\right) \Pi _{\varepsilon
_{N}}\right. \right)   \notag \\
&=&H\left( \rho _{\Pi _{\varepsilon _{N}}B}\left\Vert \sum\limits_{\alpha
_{k},N\geqslant k\geqslant 1}\prod\limits_{n=1}^{N-1}\left\vert \left\langle
\varepsilon _{n}^{\alpha _{n}}|\varepsilon _{n+1}^{\alpha
_{n+1}}\right\rangle \right\vert ^{2}\left\vert \varepsilon _{N}^{\alpha
_{N}}\right\rangle \left\langle \varepsilon _{N}^{\alpha _{N}}\right\vert
\otimes \rho _{B}^{\alpha _{1}}\right. \right)   \notag \\
&\geqslant &H\left( \rho _{\Pi _{\varepsilon _{N}}B}\left\Vert
\sum\limits_{\alpha _{k},N\geqslant k>1}\prod\limits_{n=1}^{N-1}\max_{\alpha
_{1}}\left\vert \left\langle \varepsilon _{n}^{\alpha _{n}}|\varepsilon
_{n+1}^{\alpha _{n+1}}\right\rangle \right\vert ^{2}\left\vert \varepsilon
_{N}^{\alpha _{N}}\right\rangle \left\langle \varepsilon _{N}^{\alpha
_{N}}\right\vert \otimes \rho _{B}\right. \right)   \notag \\
&=&-H(\Pi _{\varepsilon _{N}}|B)-Tr\rho _{\Pi _{\varepsilon _{N}}}\log
\sum\limits_{\alpha _{k},N\geqslant k>1}\prod\limits_{n=1}^{N-1}\max_{\alpha
_{1}}\left\vert \left\langle \varepsilon _{n}^{\alpha _{n}}|\varepsilon
_{n+1}^{\alpha _{n+1}}\right\rangle \right\vert ^{2}\left\vert \varepsilon
_{N}^{\alpha _{N}}\right\rangle \left\langle \varepsilon _{N}^{\alpha
_{N}}\right\vert \otimes \rho _{B}  \notag \\
&=&-H(\Pi _{\varepsilon _{N}}|B)-\sum\limits_{\alpha _{N}}p_{\varepsilon
_{N}^{\alpha _{N}}}\log \sum\limits_{\alpha _{k},N\geqslant k>1}\max_{\alpha
_{1}}\prod\limits_{n=1}^{N-1}\left\vert \left\langle \varepsilon
_{n}^{\alpha _{n}}|\varepsilon _{n+1}^{\alpha _{n+1}}\right\rangle
\right\vert ^{2}.
\end{eqnarray}%
The first and the second inequality is again based on the adjoint concavity
of relative entropy and the third inequality holds due to the property of
the relative entropy: $H(A||B^{\prime })\geqslant H(A||B)$, if and only if $%
B^{\prime }\geqslant B$. In order to find the tighter bound of the EUR, one
has to find the maximum of the set $\left\{ \ell _{\varepsilon }^{U}\right\}
,$ where $\ell _{\varepsilon }^{U}=-\sum\limits_{\alpha _{N}}p_{\varepsilon
_{N}^{\alpha _{N}}}\log \sum\limits_{\alpha _{k},N\geqslant k>1}\max_{\alpha
_{1}}\prod\limits_{n=1}^{N-1}\left\vert \left\langle \varepsilon
_{n}^{\alpha _{n}}|\varepsilon _{n+1}^{\alpha _{n+1}}\right\rangle
\right\vert ^{2}.$ The proof is finished.

\textsf{Proof of Eq. (\ref{totalIEP}).}From the definitions of the mutual
information $I(A$:$B)=H(\rho _{A})+H(\rho _{B})-H(\rho _{AB})$ and the
conditional entropy $H(A|B)=H(\rho _{AB})-H(\rho _{B})$, one will
immediately arrive at
\begin{equation}
H(A|B)=H(\rho _{A})-I(A:B).  \label{condi}
\end{equation}
Substitute this relation into Eq. (\ref{Main2}), we have%
\begin{eqnarray}
\sum\limits_{i=1}^{N}H(\Pi _{i}|B) &\geqslant &\mathcal{\tilde{B}}  \notag \\
&\Longrightarrow &\sum\limits_{i=1}^{N}\left[ H(\Pi _{i})-I(\Pi _{i}:B)%
\right] \geqslant \mathcal{\tilde{B}}  \notag \\
&\Longrightarrow &\sum\limits_{i=1}^{N}I(\Pi _{i}:B)\leqslant
\sum\limits_{i=1}^{N}H(\Pi _{i})-\mathcal{\tilde{B}}  \notag \\
&\Longrightarrow &\sum\limits_{i=1}^{N}I(\Pi _{i}:B)\leqslant N\log d-%
\mathcal{\tilde{B}},
\end{eqnarray}%
where the last inequality holds for $H(\Pi _{i})\leq \log d$.

\textsf{The proof of $\mathcal{\tilde{U}}_{2}$}. This proof can be done from
Eq. (\ref{z1}). Substitute Eq. (\ref{condi}) into Eq. (\ref{z1}), we arrive
at
\begin{eqnarray}
\sum\limits_{i=1}^{N}I(\Pi _{i}:B) &\leqslant &\sum\limits_{i=1}^{N}H(\Pi
_{i})+\sum\limits_{\alpha _{N}}p_{\varepsilon _{N}^{\alpha _{N}}}\log
\sum\limits_{\alpha _{k},N\geqslant k>1}\max_{\alpha
_{1}}\prod\limits_{n=1}^{N-1}\left\vert \left\langle \varepsilon
_{n}^{\alpha _{n}}|\varepsilon _{n+1}^{\alpha _{n+1}}\right\rangle
\right\vert ^{2}-(N-1)H(A|B)  \notag \\
&=&\sum\limits_{i=1}^{N-1}H(\Pi _{i})+\sum\limits_{\alpha
_{N}}p_{\varepsilon _{N}^{\alpha _{N}}}\log \frac{\sum\limits_{\alpha
_{k},N\geqslant k>1}\max\limits_{\alpha
_{1}}\prod\limits_{n=1}^{N-1}\left\vert \left\langle \varepsilon
_{n}^{\alpha _{n}}|\varepsilon _{n+1}^{\alpha _{n+1}}\right\rangle
\right\vert ^{2}}{p_{\varepsilon _{N}^{\alpha _{N}}}}-(N-1)H(A|B)  \notag \\
&\leqslant &\sum\limits_{i=1}^{N-1}H(\Pi _{i})+\log \sum\limits_{\alpha
_{k},N\geqslant k>1}\max\limits_{\alpha
_{1}}\prod\limits_{n=1}^{N-1}\left\vert \left\langle \varepsilon
_{n}^{\alpha _{n}}|\varepsilon _{n+1}^{\alpha _{n+1}}\right\rangle
\right\vert ^{2}-(N-1)H(A|B)  \notag \\
&\leqslant &(N-1)\log d+\log \sum\limits_{\alpha _{k},N\geqslant
k>1}\max\limits_{\alpha _{1}}\prod\limits_{n=1}^{N-1}\left\vert \left\langle
\varepsilon _{n}^{\alpha _{n}}|\varepsilon _{n+1}^{\alpha
_{n+1}}\right\rangle \right\vert ^{2}-(N-1)H(A|B).
\end{eqnarray}%
Here the second inequality is satisfied because of the concavity of the
logarithm function. Similarly, in order to find the tight bound of the IEP,
one has to find the minimum of the set $\left\{ {\normalsize u}_{\varepsilon
}^{I}\right\} $ with ${\normalsize u}_{\varepsilon }^{I}=\log
\sum\limits_{\alpha _{k},N\geqslant k>1}\max\limits_{\alpha
_{1}}\prod\limits_{n=1}^{N-1}\left\vert \left\langle \varepsilon
_{n}^{\alpha _{n}}|\varepsilon _{n+1}^{\alpha _{n+1}}\right\rangle
\right\vert ^{2}$.

\newpage \bigskip \noindent\textsf{Acknowledgements}

\noindent This work was supported by the National Natural Science Foundation
of China, under Grants No.11375036 and 11175033, and the Xinghai Scholar
Cultivation Plan.

\bigskip \noindent\textsf{Author contributions}

\noindent J.Z. and Y.Z. and C.-S.Y. analyzed the results and wrote the main
manuscript text. All authors reviewed the manuscript.

\bigskip \noindent\textsf{Additional Information}

\noindent Competing Financial Interests: The authors declare no competing
financial interests.


\begin{thebibliography}{99}
\bibitem{HUP} Heisenberg, W. J. Z. \"{U}ber den anschaulichen Inhalt der
quantentheoretischen Kinematik und Mechanik. \textit{Z. Phys.} \textbf{43},
172 (1927).

\bibitem{kennard} Kennard, E. H. Zur Quantenmechanik einfacher
Bewegungstypen. \textit{Z. Phys.} \textbf{44}, 326 (1927).

\bibitem{RHUP} Robertson, H. P. The Uncertainty Principle. \textit{Phys. Rev.%
} \textbf{34}, 163 (1929).

\bibitem{Wehner} Wehner, S. \& Winter, A. Entropic uncertainty relations-a
survey. \textit{New J. Phys.} \textbf{12}, 025009 (2010).

\bibitem{inade} Bialynicki-Birula, I. \& Rudnicki, L. Statistical
Complexity. edited by K. D. Sen, (Springer, New York, 2011).

\bibitem{Ozawa0} Ozawa, M. Universally valid reformulation of the Heisenberg
uncertainty principle on noise and disturbance in measurement. \textit{Phys.
Rev. A} \textbf{67}, 042105 (2003).

\bibitem{Werner} Busch, P., Lahti, P. \& Werner, R. F. Proof of Heisenberg's
Error-Disturbance Relation. \textit{Phys. Rev. Lett.} \textbf{111}, 160405
(2013).

\bibitem{Wernerz} Busch, P., Lahti, P. \& Werner, R. F. Quantum
root-mean-square error and measurement uncertainty relations. \textit{Rev.
Mod. Phys.} \textbf{86}, 1261 (2014).

\bibitem{shiyan} Erhart, J. et al. Experimental demonstration of a
universally valid error--disturbance uncertainty relation in spin
measurements. \textit{Nat. Phys.} \textbf{8}, 185 (2012).

\bibitem{shiyan1} Kaneda, F., Baek, S. -Y., Ozawa, M., \& Edamatsu, K.
Experimental Test of Error-Disturbance Uncertainty Relations by Weak
Measurement \textit{Phys. Rev. Lett.} \textbf{112}, 020402 (2014).

\bibitem{shiyan2} Rozema, L. A. et al. Violation of Heisenberg's
Measurement-Disturbance Relationship by Weak Measurements. \textit{Phys.
Rev. Lett.} \textbf{109}, 100404 (2012).

\bibitem{shiyan4} Baek, S. Y., Kaneda, F., Ozawa, M. \& Edamatsu,K.
Experimental violation and reformulation of the Heisenberg's
error-disturbance uncertainty relation. \textit{Sci. Rep.} \textbf{3}, 2221
(2013).

\bibitem{shiyan5} Sulyok, G. et al. Violation of Heisenberg's
error-disturbance uncertainty relation in neutron-spin measurements. \textit{%
Phys. Rev. A} \textbf{88}, 022110 (2013).

\bibitem{shiyan6} Ringbauer, M. et al. Experimental Joint Quantum
Measurements with Minimum Uncertainty. \textit{Phys. Rev. Lett.} \textbf{112}%
, 020401 (2014).

\bibitem{Deutsch} Deutsch, D. Uncertainty in Quantum Measurements \textit{%
Phys. Rev. Lett.} \textbf{50}, 631 (1983).

\bibitem{1975shannon} Bialynicki-Birula, I. \& Mycielski, J. Uncertainty
Relations for Information Entropy in Wave Mechanics. \textit{Commun. Math.
Phys.} \textbf{44}, 129 (1975).

\bibitem{Uffink} Maassen, H. \& Uffink, J. B. M. Generalized Entropic
Uncertainty Relations. \textit{Phy. Rev. Lett.} \textbf{60}, 1103 (1988).

\bibitem{Hall} Hall, M. J. W. Information Exclusion Principle for
Complementary Observables. \textit{Phys. Rev. Lett.} \textbf{74}, 3307
(1995).

\bibitem{Grudka} Grudka, A. et al. Conjectured strong
complementary-correlations tradeoff. \textit{Phys. Rev.A} \textbf{88},
032106 (2013).

\bibitem{Piani} Coles, P. J. \& Piani, M. Improved entropic uncertainty
relations and information exclusion relations. \textit{Phys. Rev. A} \textbf{%
89}, 022112 (2014).

\bibitem{Bolan} Rudnicki, \L . Z., Pucha\l a, \& \'{Z}yczkowski,K. Strong
majorization entropic uncertainty relations. \textit{Phys. Rev. A} \textbf{89%
}, 052115 (2014).

\bibitem{Heisenberg} Busch, P., Heinonen, T., \& Lahti, P. Heisenberg's
uncertainty principle. \textit{Phys. Reports} \textbf{452}, 155 (2007).

\bibitem{Ozawa1} Buscemi, F., Hall, M. J. W., Ozawa, M. \& Wilde, M. M.
Noise and Disturbance in Quantum Measurements: An Information-Theoretic
Approach. \textit{Phys. Rev. Lett.} \textbf{112}, 050401 (2014).

\bibitem{nature} Berta, M. et al. The uncertainty principle in the presence
of quantum memory. \textit{Nat. Phys.} \textbf{6}, 659 (2010).

\bibitem{science} Oppenheim, J. \& Wehner, S. The Uncertainty Principle
Determines the Nonlocality of Quantum Mechanics. \textit{Science} \textbf{330%
}, 1072 (2010).

\bibitem{Heisen} Busch, P., Lahti, P. \& Werner, R. F. Heisenberg
uncertainty for qubit measurements. \textit{Phys. Rev. A} \textbf{89},
012129 (2014).

\bibitem{nature1} Prevedel, R. et al. Experimental investigation of the
uncertainty principle in the presence of quantum memory and its application
to witnessing entanglement. \textit{Nat. Phys.} \textbf{7}, 757 (2011).

\bibitem{wit} Li, C.-F. et al. Experimental investigation of the
entanglement-assisted entropic uncertainty principle. \textit{Nat. Phys.}
\textbf{7}, 752 (2011).

\bibitem{miyao} Tomamichel, M. \& Renner, R. Uncertainty Relation for Smooth
Entropies. \textit{Phys. Rev. Lett.} \textbf{106}, 110506 (2011).

\bibitem{miyao1} Tomamichel, M., Lim, C. C. W., Gisin, N. \& Renner, R.
Tight finite-key analysis for quantum cryptography. \textit{Nat. Commun.}
\textbf{3}, 634 (2012).

\bibitem{Spiros} Kechrimparis, S. \& Weigert, S. Heisenberg uncertainty
relation for three canonical observables. \textit{Phys. Rev. A} \textbf{90},
062118 (2014).

\bibitem{Liu} Liu, S., Mu, L.-Z., \& Fan, H. Entropic uncertainty relations
for multiple measurements. \textit{arXive:} 1410.5177.

\bibitem{nielsen} Nielsen M. A. \& Chuang I. L., Quantum computation
and Quantum information. Theorem 11.8 on Page 513 (Cambridge
University Press, Cambridge, 2010).

\bibitem{wernerstate} Werner, R. F. Quantum states with
Einstein-Podolsky-Rosen correlations admitting a hidden-variable model.
\textit{Phys. Rev. A} \textbf{40}, 4277 (1989).

\bibitem{horodec} Horodecki, P. Separability criterion and inseparable mixed
states with positive partial transposition. \textit{Phys. Letts. A} \textbf{%
232}, 333 (1997).
\end{thebibliography}
\end{document}